\documentclass[sigconf,table,xcdraw,screen]{acmart}
\usepackage[ruled,vlined]{algorithm2e}
\usepackage{cellspace}
\usepackage{graphicx}
\usepackage{paralist}
\usepackage{tabularx}
\usepackage{balance}
\usepackage{multirow}
\usepackage{multicol}
\usepackage{pbox}
\usepackage{mathtools}
\usepackage[caption=false]{subfig}
\usepackage{algorithmic}
\usepackage{paralist}
\usepackage{url}
\usepackage{hyperref}
\usepackage{amsmath}
\usepackage{setspace}
\usepackage{verbatim}
\usepackage{float}
\usepackage[normalem]{ulem}
\usepackage{xspace}
\usepackage{ifthen}
\usepackage{pifont}
\usepackage{tikz}
\usepackage{array}

\usepackage{hyperref}
\include{macro}

\newcommand{\Comet}{{\sc Comet}\xspace}
\newcommand{\Comets}{{\sc Comet's}\xspace}
\newcolumntype{C}[1]{>{\centering\arraybackslash}p{#1}}

\newcommand{\secref}[1]{Sec.~\ref{#1}\xspace}

\newcommand{\figref}[1]{Fig.~\ref{#1}\xspace}

\newcommand{\tabref}[1]{Table~\ref{#1}\xspace}

\copyrightyear{2020} 
\acmYear{2020} 
\setcopyright{acmlicensed}
\acmConference[ICSE '20]{42nd International Conference on Software Engineering}{May 23--29, 2020}{Seoul, Republic of Korea}
\acmBooktitle{42nd International Conference on Software Engineering (ICSE '20), May 23--29, 2020, Seoul, Republic of Korea}
\acmPrice{15.00}
\acmDOI{10.1145/3377811.3380418}
\acmISBN{978-1-4503-7121-6/20/05}

\begin{document}

\title{Improving the Effectiveness of Traceability Link Recovery \\ using Hierarchical Bayesian Networks}

\author{Kevin Moran}
\affiliation{%
  \institution{William \& Mary}
  \city{Williamsburg} 
  \state{VA} 
  \country{USA}
}
\email{kpmoran@cs.wm.edu}

\author{David N. Palacio}
\affiliation{%
  \institution{William \& Mary}
  \city{Williamsburg} 
  \state{VA} 
  \country{USA}
}
\email{dnaderp@cs.wm.edu}

\author{\mbox{Carlos Bernal-C\'{a}rdenas}}
\affiliation{%
  \institution{William \& Mary}
  \city{Williamsburg} 
  \state{VA} 
  \country{USA}
}
\email{cebernal@cs.wm.edu}

\author{Daniel McCrystal}
\affiliation{%
  \institution{William \& Mary}
  \city{Williamsburg} 
  \state{VA} 
  \country{USA}
}
\email{dmc@cs.wm.edu}

\author{Denys Poshyvanyk}
\affiliation{%
  \institution{William \& Mary}
  \city{Williamsburg} 
  \state{VA} 
  \country{USA}
}
\email{denys@cs.wm.edu}

\author{Chris Shenefiel}
\affiliation{%
  \institution{Cisco Advanced Security Research Group}
  \city{Morrisville} 
  \state{NC} 
  \country{USA}
}
\email{cshenefi@cisco.com}

\author{Jeff Johnson}
\affiliation{%
  \institution{Cisco Security and Trust Engineering}
  \city{Morrisville} 
  \state{NC} 
  \country{USA}
}
\email{johnsonj@cisco.com}

\renewcommand{\shortauthors}{K. Moran, D. N. Palacio, C. Bernal-C\'{a}rdenas, D. McCrsytal, \\D. Poshyvanyk, C. Shenefiel, \& J. Johnson}

\begin{abstract}

Traceability is a fundamental component of the modern software development process that helps to ensure properly functioning, secure programs. Due to the high cost of manually establishing trace links, researchers have developed automated approaches that draw relationships between pairs of textual software artifacts using similarity measures. However, the effectiveness of such techniques are often limited as they only utilize a single measure of artifact similarity and cannot simultaneously model (implicit and explicit) relationships across groups of diverse development artifacts.

In this paper, we illustrate how these limitations can be overcome through the use of a tailored \textit{probabilistic model}.  To this end, we design and implement a Hierarchi\textbf{C}al Pr\textbf{O}babilistic \textbf{M}odel for Softwar\textbf{E} \textbf{T}raceability (\Comet) that is able to infer candidate trace links. \Comet is capable of modeling relationships between artifacts by combining the complementary observational prowess of multiple measures of textual similarity. Additionally, our model can holistically incorporate information from a diverse set of sources, including developer feedback and transitive (often implicit) relationships among groups of software artifacts, to improve inference accuracy. We conduct a comprehensive empirical evaluation of \Comet that illustrates an improvement over a set of \textit{optimally configured} baselines of $\approx$14\% in the best case and $\approx$5\% across all subjects in terms of average precision. The comparative effectiveness of \Comet \textit{in practice}, where optimal configuration is typically not possible, is likely to be higher. Finally, we illustrate \Comets potential for practical applicability in a survey with developers from Cisco Systems who used a prototype \Comet Jenkins plugin.
\vspace{-0.2cm}
\end{abstract}



 % NOT required for Proceedings

\maketitle

%-----------------------------------
\section{Introduction \& Motivation}
\label{sec:intro}

The importance of traceability in modern software systems cannot be overstated. Traceability links that connect ``high-level" artifacts such as requirements and use cases to ``low-level'' artifacts written in code help to facilitate crucial components of the software development and maintenance cycle. For instance, linking requirements to code provides visibility into a system by enumerating what has been implemented, whereas linking requirements to test cases helps to provide an indication that the software is functioning as expected. Additionally, the establishment of trace links aids in facilitating a broad set of developer activities including code comprehension, change impact analysis, and compliance validation~\cite{Cleland-Huang:Springer'12}.  In certain software domains, such as those involving safety critical systems, traceability is necessarily \textit{mandated} by regulatory bodies in order to properly demonstrate the safe functioning of a system~\cite{Nejati:IST'12,Rempel:ICSE'14,Cleland-Huang:ICSE'10,Mader:Soft'13}. Furthermore, traceability is increasingly used to help ensure the \textit{security} of a given system~\cite{Nhlabatsi:SST'15}. For example, our industrial partners at Cisco Systems, Inc. require that security-critical requirements are verified by a dedicated group of analysts to avoid software threats and ensure best practices. 

Unfortunately, despite its importance, software traceability is, by its nature, an inherently difficult and error prone task~\cite{Cleland-Huang:FOSE'14,Mahmoud:ICPC'12,Mader:Soft'13}. This difficulty primarily stems from the need to bridge a logical abstraction gap that exists between different software artifacts, such as requirements written in natural language and code written in ``lower-level'' programming languages. Given the effort required to establish and evolve effective trace links, it is often too costly to manually establish them outside of regulated domains, and in practice the quality of mandated links are often questionable~\cite{Cleland-Huang:FSE'14}.  

The inherent difficulty in establishing trace links has lead to research on automated techniques for modeling, establishing, and evolving trace links that primarily rely upon information retrieval (IR)~\cite{Lucia:ICSM'04,Dekhtyar:RE'07,Asuncion:ICSE'10,McMillan:TEFSE'09,Gethers:ICSM'11,DeLucia:ASE'08,DeLucia:EMSE'09,Mahmoud:ICPC'12,Antoniol:ICSE'00,Marcus:ICSE'03,Mills:ICSME18,Jiang:ASE'08,Kuang:SANER'17} and machine learning (ML)~\cite{Mahmoud:RE'16,Guo:MSR'16,Asuncion:ICSE'10,Spanoudakis:SEKE'03,Falessi:EMSE17} techniques which retrieve or predict trace links based upon textual similarity metrics.  However, in large part, current automated approaches for traceability often trade precision for completeness and vice versa, making them difficult to adopt in practice. We observe three major shortcomings of current automated techniques that contribute to their limited effectiveness:

\noindent{\textbf{1) Limited Measures of Artifact Similarity:}} Existing techniques for trace link recovery tend to use a single textual similarity metric to draw relationships between artifacts. This is problematic for several reasons. Perhaps most importantly, it is often difficult or impossible to determine how well a technique that uses a given similarity measure will function on artifacts from a new project without any pre-existing trace links. This so-called ``cold-start'' problem is due to the fact that existing IR/ML techniques for measuring textual similarity often need to be calibrated on a subset of "ground-truth" artifact pairs with pre-existing links. This makes performance of these techniques difficult to predict when applied to new datasets. Furthermore, in practice industrial projects often lack pre-existing trace links, as confirmed by our  partners at Cisco. Thus, while certain techniques have been shown to perform well on research benchmarks, the efficacy of a similarity measure is often tightly coupled to the underlying semantics of software artifact text~\cite{Lohar:FSE'13,Guo:ICSE'17,Biggerstaff:ACM'94}, and to the configuration of the corresponding IR/ML technique~\cite{Oliveto:ICPC'10}. 

Using only a single textual similarity metric also needlessly restricts the predictive power of a traceability technique. Past work has illustrated the orthogonality of different similarity measures~\cite{Oliveto:ICPC'10}, suggesting that combining \textit{several} different measures could lead to more accurate and robust techniques that function \textit{consistently} well when applied to new projects without pre-existing links.

\noindent{\textbf{2) Inability to Effectively Capture Developer Feedback:}} The rapid pace of modern agile development practices often results in crucial knowledge about a software system being siloed within the expertise of individual developers. Thus, one unstructured development artifact that has gone underutilized by past techniques is \textit{developer feedback}. When an automated traceability model is uncertain about particular trace link pairs, developers can provide critical feedback to help improve trace link inference.

\noindent{\textbf{3) Limited View of Interactions Between Artifacts}}
Existing automated traceability approaches are typically tailored to establish relationships between pairs of specific types of artifacts (\eg user stories and class files). However, information pertaining to the relationship of one type of artifact pair may be contained within other related artifacts. For example, if a piece of source code is linked to a given requirement through textual similarities, and this source code is also intrinsically linked to test code via method calls, then it is likely the requirement is also linked to the test code. However, in this situation, it may be difficult for a textual similarity metric to link the requirements and test code, due to limited test documentation, for example. Thus, in this way, established relationships between certain artifacts may influence the probability of other artifact relationships. In this paper we refer to these phenomena as \textit{transitive links}. Existing techniques generally cannot model such interactions between artifacts.

The limitations discussed above stem from both technical and practical limitations of existing traceability techniques, and surfaced during our development of an automated traceability approach in close collaboration with Cisco Systems. In this paper we introduce a novel technique that overcomes these limitations by constructing a Hierarchical Bayesian Network for inferring a set of candidate trace links. The model that underlies our approach is capable of deriving the probability that a trace link exists between two given artifacts by combining information from multiple measures of textual similarity, while simultaneously modeling transitive relationships and accounting for developer expertise.  We implemented our approach, called \Comet (Hierarchi\textbf{C}al Pr\textbf{O}babilistic \textbf{M}odel for Softwar\textbf{E} \textbf{T}raceability), in both an extensible Python library and as a plugin for the popular Jenkins CI/CD system.  In an extensive set of empirical experiments we illustrate that \Comet is able to outperform the median precision of \textit{optimally configured} baseline techniques by $\approx$5\% across subjects and $\approx$14\% in the best case. Given that optimal configuration is typically not possible in practice, this illustrates that given a project with no pre-existing trace links, \Comet is likely to perform significantly better than most existing IR/ML techniques. Additionally, we show \Comets potential for integration into the workflows of development teams at Cisco. In summary, this paper's contributions are as follows:

\vspace{-0.1cm}
\begin{itemize}
	\item{The derivation of a Hierarchical Bayesian Network (HBN) for inferring a candidate set of trace links;}	
	\item{An implementation of this model, called \Comet, as both an extensible Python library and a Jenkins plugin that has been deployed for testing with our industrial partners at Cisco;}
	\item{An extensive evaluation of \Comet on both open source projects and two industrial datasets from one industrial software project, including feedback from professional developers at a major telecommunication software company;}
	\item{An open source, commercial-grade traceability benchmark, developed in coordination with our industrial partner, for the benefit of the research community;}
	\item{An online appendix, including our open source implementation of \Comet and evaluation data for reproducibility~\cite{appendix}}.
\end{itemize}

%\vspace{-0.3cm}
\section{Related Work}
\label{sec:related-work}
%\vspace{-0.2cm}

We focus our discussion of related work on prior techniques that have, in limited contexts, (i) considered novel or hybird textual similarity measures, (ii) modeled the effects of multiple types of artifacts, or (iii) incorporated developer expertise. We then conclude with a statement distilling \Comets novelty.

\noindent{\textbf{Novel/Hybird Textual Similarity Measures:}} Guo \etal~\cite{Guo:ICSE'17} proposed an approach for candidate trace link prediction that uses a semantically enhanced similarity measure based on Deep Learning (DL) techniques. However, unlike \Comet, this technique requires pre-existing trace links in order to train the DL classifier.  In contrast, \Comet does not require known links for the projects it is applied to, but rather requires a project to serve as a tuning set. We show that \Comet performs well when tuned and tested on different datasets, outperforming Guo \etals DL-based approach when it is trained in a similar manner. Gethers \etal~\cite{Gethers:ICSM'11}, implemented an approach that is capable of combining information from canonical IR techniques (\ie VSM, Jensen-Shannon) with Topic Modeling techniques. However, their approach can only combine two IR/ML techniques, whereas \Comet can combine and leverage the observations from several IR/ML techniques, and combine this with other information such as expert feedback and transitive links. 

\noindent{\textbf{Modeling of Multiple Artifacts:}} Rath \etal~\cite{Rath:ICSE'18} recently explored linking nontraditional information including issues and commits, and Cleland-Huang \etal~\cite{Cleland-Huang:ICSE'10} have investigated linking regulatory codes to product level requirements.  \Comets model has the potential to improve trace link recovery in these scenarios both through its more robust modeling of textual similarity, and through incorporation of transitive link information. Furtado \etal~\cite{Furtado:RE'16}, explored traceability in the context of agile development, and Nishikawa \etal~\cite{Nishikawa:ICSME'15} first explored the use of transitive links in a deterministic traceability model. Additionally, Kuang \etal used the closeness of code dependencies, to help improve IR-based traceability recovery~\cite{Kuang:SANER'17}. However, none of these approaches is capable of incorporating transitive links while also considering combined textual similarity metrics and developer feedback.

\noindent{\textbf{Incorporation of Developer Expertise:}} De Lucia \etal \cite{DeLucia:ICSM'06} and Hayes \etal~\cite{Hayes:TSE'06} analyzed approaches that use relevance feedback to improve trace link recovery. However, these approaches are either tied to a particular type of model (such as TF-IDF~\cite{DeLucia:ICSM'06}), or require knowledge of the underlying model to function optimally. In contrast, \Comet implements a lightweight, likert-based feedback collection mechanism that we illustrate can improve link accuracy even when only a small amount of feedback is collected.
 
\noindent{\textbf{Summary of Advancement over Prior Work:}} \Comets features facilitate its application to projects without any pre-existing trace links, and as our evaluation illustrates, allow it to perform consistently well across datasets. \Comet is able to combine information from transitive links with both robust textual similarity measures and lightweight developer feedback for improved accuracy. While some aspects of \Comets approach have been considered in limited contexts in prior work -- such as developer feedback~\cite{DeLucia:ICSM'06,Hayes:TSE'06} and restricted combinations of IR/ML techniques~\cite{Gethers:ICSM'11} -- there has never been a framework capable of combining all these aspects in a holistic approach. Our evaluation illustrates that \Comets holistic HBN is able to outperform baseline techniques on average.

%\vspace{-0.3cm}
\section{Background}
\label{sec:background}

%-------------------------------
\subsection{Problem Definition}
\label{subsec:traceability-background}

Our goal is to design a model that captures meaningful information regarding logical relationships between software artifacts, and then use this model to infer a set of candidate trace links. More specifically, given a set of source artifacts $S$ (\eg requirements, use cases) such that $S = \{S_{1},S_{2},\ldots S_{n}\}$ and a set of target artifacts $T$ (\eg source code files, test cases) such that $T = \{T_{1},T_{2},\ldots T_{n}\}$, we aim to infer whether a trace link $L$ exists between all possible pairs of artifacts in $S$ and $T$ such that $L = \{(s,t) | s\in S, t\in T, s\leftrightarrow t\}$ where each pair of artifacts $s$ and $t$ are said to be logical trace links.

%-------------------------------
\vspace{-0.2cm}
\subsection{\hspace{-0.1cm}Defining a Probabilistic View of Traceability}
\label{subsec:model-motivation}

\subsubsection{\textbf{The Probabilistic Nature of Software Traceability}} 

The process of building software is not inherently deterministic, and is instead the result of decisions made by engineers over prolonged periods of time that may be hard to predict. Developer decisions related to nearly every observable phenomenon in modern software development are influenced by a combination of multiple factors. For instance, the presence of a functional bug may be influenced by the quality of related requirements, implementation constraints imposed by a given programming language~\cite{Ray:CACM'17}, or the change-proneness of underlying APIs~\cite{Linares-Vasquez:FSE'13}. Given that such factors are often hard to predict, there is a clear sense of randomness inherent to the software development process. Similarly, the existence of trace links among software artifacts is also likely to be influenced by several different effectively \textit{random} factors.  

These factors could include textual similarities between artifacts, programmatic associations between pieces of code, or even abstract notions of similarity held by expert developers.  For example, the textual quality of requirements or identifiers in code are typically a function of several factors such as the fluency and writing style of the author and the familiarity of key phrases chosen for identifiers \cite{Dasgupta:ICSME'13}. This may lead to variable names that may be perfectly clear to one engineer being indecipherable to another.  \textit{From this view point, the existence of trace links between software artifacts can be thought of as an inherently a probabilistic phenomenon}.

\subsubsection{\textbf{Traceability as a Bayesian Inference Problem}} 

Hence, in order to effectively model trace links among software artifacts, it is necessary to model a collection of random factors that influence \textit{the probability that a trace link exists}. Thus, the process of deriving trace links can be modeled as a \textit{bayesian inference} problem, wherein a probability distribution representing the existence of a trace link between two artifacts can be inferred. As we illustrate, by modeling the trace link recovery problem in a probabilistic manner, we are able to construct an an automated approach that largely overcomes the typical drawbacks discussed in Sec.~\ref{sec:intro}. To understand this context, let us consider the general definition of Bayes' Theorem:

\vspace{-0.27cm}
\begin{equation}
P(H|O) = \frac{P(O|H)\cdot P(H)}{P(O)}
\end{equation}
\vspace{-0.27cm}

\noindent where $H$ is a hypothesis regarding some phenomenon, $O$ is a set of observations that provide some information about the hypothesis, and where our goal is to infer or estimate the probability that our hypothesis is true $P(H|O)$, which is called the \textit{posterior probability distribution}, or more simply the \textit{posterior}. However, a given hypothesis is rarely made in a vacuum, and one typically holds some \textit{prior belief} as to the probability that is being inferred.  This prior belief is modeled as a probability distribution $P(H)$, which we will simply refer to as the \textit{prior}, and can be influenced by a number of factors. In order for the posterior to be inferred from a set of observations, these must be modeled in a probabilistic manner. This is the purpose of the \textit{likelihood} $P(O|H)$, which is a probability distribution that is derived purely from observed data. Thus, in Bayesian inference initial beliefs are represented as the prior, observations are modeled as the likelihood and the final beliefs are represented by the posterior. This posterior probability distribution can be \textit{inferred} via one of several existing statistical inference techniques. In framing the problem of inferring trace links as a Bayesian problem, we consider our hypothesis to be whether a given trace link exists between a single source artifact $S_x$ and a single target artifact $T_y$. Given the nature of trace links (\eg a link either does or does not exist) we can model our prior as a distribution on the interval $[0,1]$, where 1 indicates the presence of a link and 0 indicates an absence. 

\subsubsection{\textbf{A Hierarchical Bayesian Network for Traceability}} 

In the context of this paper, we will consider our \textit{likelihood} (observations) to be the binary indication that a link exists according to a set of textual similarity measures and an empirically derived threshold value. However, given that we aim to model multiple factors that might influence traceability, our model employs multiple \textit{priors}, called \textit{hyperpriors}, forming a Hierarchical Bayesian Network (HBN). In this work, we consider three priors corresponding to the three factors we wish to model: (i) a normalized set of diverse textual similarity measures, (ii) developer expertise, and (iii) transitive trace links. We assign each of these priors an initial probability distribution, which is then influenced and estimated based upon observable data (e.g. a developer confirming or denying a trace link). Once this network is established, the \textit{posterior} can be computed via one of several estimation techniques. By modeling these three information sources, our technique is able to largely overcome the limitations enumerated in Sec.~\ref{sec:intro}. HBNs are also highly extensible via adjustments to the prior(s). Thus our defined model be capable of adapting to advancements in textual similarity measures or considering new development artifacts from future development workflows.

\vspace{-0.2cm}
\section{Inferring Trace Links with a Hierarchical Bayesian Network}
\label{sec:approach}

\begin{figure}[tb]
\centering
\vspace{-0.1cm}
\includegraphics[width=0.95\columnwidth]{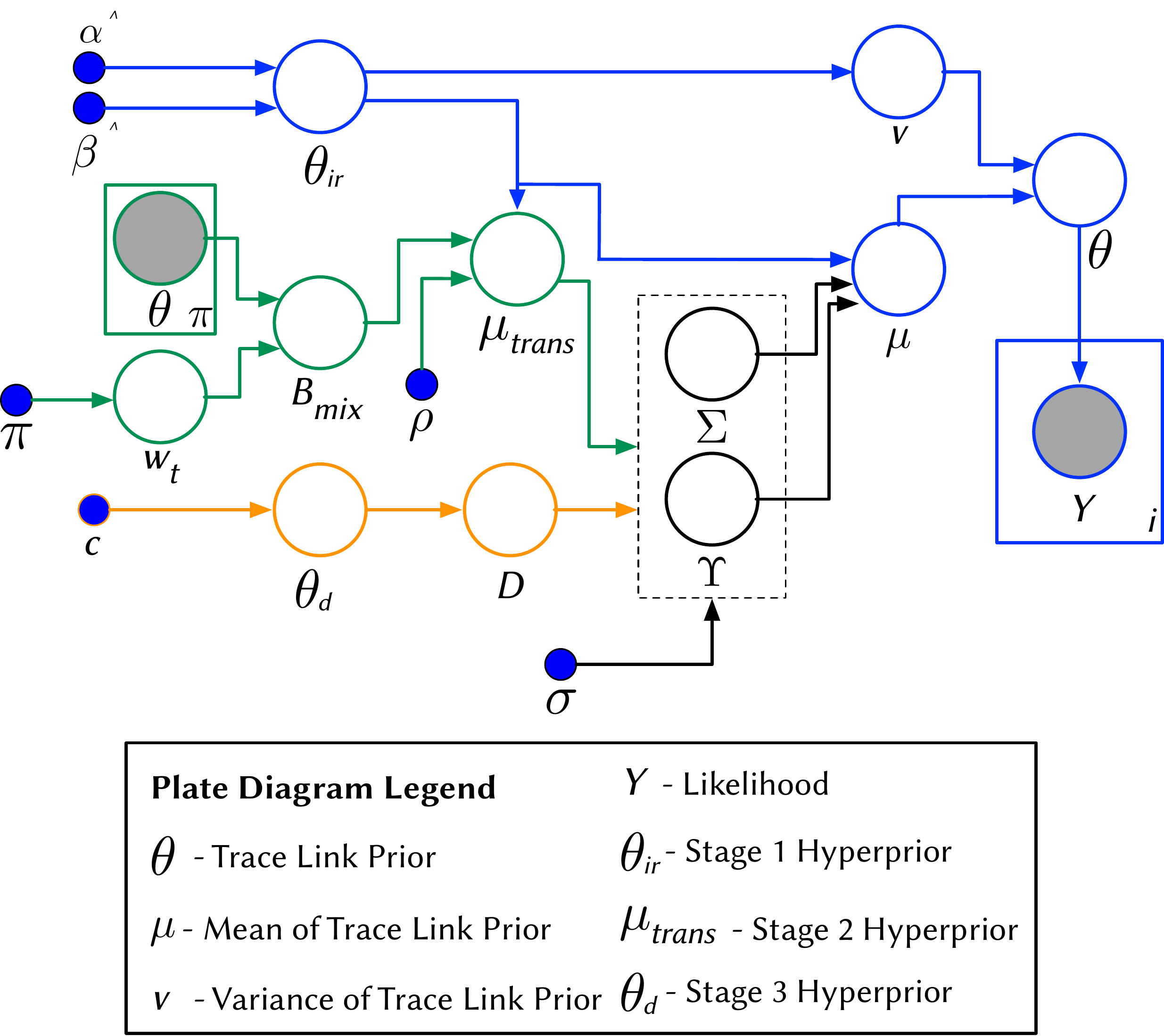}
\vspace{-0.2cm}
\caption{Plate Diagram of \Comets HBN}
\label{fig:model}
\vspace{0.0cm}
\end{figure}

In this section, we provide a formal description of \Comets probabilistic model introduced at a high level in Sec. \ref{subsec:model-motivation}. To help aid in the comprehension of \Comets underlying model, we provide a graphical representation using plate notation~\cite{Murphy:2012} in Fig. \ref{fig:model}, which we use to guide our introduction and discussion. The model in Fig. \ref{fig:model} is computed on a \textit{per link} basis, that is between all potential links between a set of source ($S$) and target artifacts ($T$). In this section we will use $S_x$ and $T_y$ to refer to a single source and target artifact of interest respectively. \Comets probabilistic model is formally structured as an HBN, centered upon a \textit{trace link prior} $\theta$ which represents the model's prior belief about the probability that $S_x$ and $T_y$ are linked.  Our model is hierarchical, as the trace link prior is influenced by a number of \textit{hyperpriors}, which are constructed in accordance with a set of \textit{hyper-parameters} that are either derived empirically, or fixed. In Fig.~\ref{fig:model}, hyperpriors are represented as empty nodes, and hyper-parameters are represented as shaded blue nodes. In general, empty nodes represent latent, or hidden, variables whereas shaded nodes represent variables that are known or empirically observed quantities. The rectangles, or ``plates'' in the diagram are used to group together variables that repeat.

To make our model easier to comprehend, we have broken it down into four major configurations, which we call \textit{stages}, indicated by different colors in Fig. \ref{fig:model}. The first stage of our model (shown in blue at top) unifies the collective knowledge of textual similarity metrics computed by IR/ML techniques. The second stage (shown in orange at the bottom) reconciles expert feedback to improve the accuracy of inferred trace links. The third stage (shown in green in the middle) accounts for transitive relationships among development artifacts, and the fourth stage combines each of the underlying stages. It should be noted that the first stage of our model can be taken as the ``base case'' upon which the other complexities build and is always required to infer the existence of a trace link.  The order of calculation starts with the first stage and proceeds sequentially. The design and parameterization of our model presented in this section is not arbitrary, but instead based on the well-founded theory of \textit{conjugate priors}~\cite{Raiffa:61} which aids in defining appropriate distributions and hyperparameters for a given prior. We center the description of our model first upon the likelihood estimation and then around the estimation of the prior probability distribution as defined by the four stages. After defining the hyperpriors for each of the four stages we briefly discuss the inference techniques we employ to estimate the posterior probability distribution of our model and thus the probability of whether a given link exists. While this section provides an overview of our model, we discuss its instantiation (including utilized IR/ML techniques) in Sec. \ref{subsec:exp-context}.

\subsection{Estimating the Likelihood}
\label{subsubsec:model-likelihood}

The likelihood function in our HBN models observed data so that these observations can be reconciled with our estimated prior probability distribution to infer a posterior probability.  The likelihood is shown as the observed variable $Y$ (Fig. \ref{fig:model}). The variable $i$ represents the number of observations made. In the context of traceability, we express the likelihood as a discrete Bernoulli distribution, as two artifacts can either be ``linked'' or ``not-linked'':

\vspace{-0.3cm}
\begin{equation}\label{eq:likelihood}
Y = p(l_i|\theta_i)= Bern(l_i|\theta_i)
\end{equation}

\noindent where $l_i$ is an observable data point ${0,1}$ for $i$ number of observations. We define an observation as a function of the textual similarity score generated by an IR technique between $S_x$ and $T_y$ and some threshold $k_i$ where any similarity above the threshold is considered an observed link, and any similarity value below this threshold is considered a non-link. The number of IR techniques or configurations utilized corresponds to the number of observations $i$.  Ideally, to capture the most accurate trace link observations from IR techniques, the threshold $k_i$ should be chosen to maximize the chance that each IR technique correctly establishes whether two artifacts are linked.  In other words, $k_i$ should be chosen for each IR technique such that the precision and recall of the technique is maximal across the entire set of considered source and target artifacts $S$ and $T$.  However, this information is not available a-priori without the consultation of a ground truth set of trace links. As we illustrate in \secref{sec:study} this threshold can often be estimated with surprising accuracy by analyzing the distribution of similarity values an IR technique produces for a given set of artifacts.

\subsection{Stage 1 - Unifying Textual Similarities between Development Artifacts}
\label{subsec:model-comp1}

	The first ``base'' stage of our model informs the trace link prior, represented as a probability distribution $p(\theta)$, according to the textual similarity measurements of a set of IR techniques.  However, converse to the likelihood estimation, the actual textual similarity values of IR techniques are directly used to estimate a Beta distribution (the conjugate prior of the likelihood's Bernoulli distribution). This Beta distribution is represented as follows: 
\vspace{-0.2cm}
\begin{equation}\label{eq:lvl1-dist}
\theta \sim B(\mu, \nu) 
\end{equation}

\noindent where $\mu$ and $\nu$ are parameters of the Beta distribution representing its mean and variance. This prior, and its two parameters are illustrated in the right-most part of the blue segment (Fig. \ref{fig:model}). To inform this Beta distribution, the textual similarity of values of a given number $i$ of IR/ML techniques are normalized according to a sigmoid function centered upon the median of the distribution of similarity values across all $S$ and $T$ in a given dataset. Then a logistic regression is performed upon the normalized similarity values to infer the values $\hat{\alpha}$ and $\hat{\beta}$ which define a hyperprior beta distribution $\theta_{IR}$. This hyperprior with parameters are shown on the left of the blue segment (Fig. \ref{fig:model}). The mean and the variance of this hyperprior distribution then inform $\mu$ and $\nu$ of the base prior $\theta$:

\vspace{-0.3cm}
\begin{equation}\label{eq:lvl1-params}
\nu = Var[\theta_{IR}] \quad \mu = Mean[\theta_{IR}]
\end{equation}

\noindent Thus, by considering the textual similarity values of a set of IR techniques, our model can effectively reconcile the collective knowledge to ultimately make an informed prediction.

\subsection{Stage 2 - Incorporating Developer Feedback}
\label{subsec:model-comp2}

	The second stage of our model is capable of leveraging human feedback by influencing the prior distribution introduced in the first stage of our model. To model expert feedback, we estimate hyperpriors $D$ and $\theta_d$, shown in orange in Fig. \ref{fig:model}. To perform this estimation, our model accepts from a developer or analyst, their confidence that a given link exists as a value between $[0,1]$. In Section \ref{subsec:comet-jenkins} we illustrate how such feedback can be collected from developers in a lightweight manner. This confidence value serves as a parameter for estimating the distribution of the first hyperprior:

\vspace{-0.3cm}
\begin{equation}
\theta_{d} \sim B(\mu_{d}=c,sd=0.01)	
\end{equation}

\noindent where $\theta_{d}$ is a Beta distribution parameterized by its mean $\mu_d$ set to the confidence value provided by a developer, and standard deviation $sd$ which we set to 0.01 signaling a low variance in the derived Beta distribution. This distribution then parameterizes the second hyperprior $D$, modeled as a Bernoulli distribution. Now that we have derived a distribution representing developer feedback, we must define how this distribution affects the prior probability of the first stage of our model. To do this, we define reward and penalty functions $\Upsilon$ and $\Sigma$ that are influenced by $\sigma$ which represents a specified \textit{belief factor} between $[0,1]$ that controls the extent to which the feedback influences the trace link prior. The reward function is defined as $\Upsilon = \sigma*D$, whereas the penalty function is defined as $\Sigma = \sigma*(D-1)$. These factors impact the first stage prior Beta distribution by affecting its mean $\mu$:

\vspace{-0.3cm}
\begin{equation}\label{eq:mean-affect}
	\mu \sim N(\mu_n= \Sigma + \Upsilon, sd=0.01)
\end{equation}

\noindent where the mean of the first stage prior is represented as a normal distribution parameterized by $\mu_n$ set to the sum of $\Sigma$ and $\Upsilon$, and a standard deviation set to 0.01. Thus, in this manner, expert feedback is utilized to influence the prior distribution that a given trace link exists. The structure of the Stage 2 hyperpriors allows \Comets HBN to effectively consider feedback from multiple developers.

\subsection{Stage 3 - Leveraging Transitive Links}
\label{subsec:model-comp3}

\begin{figure}[tb]
\centering
\vspace{-0.35cm}
\includegraphics[width=\columnwidth]{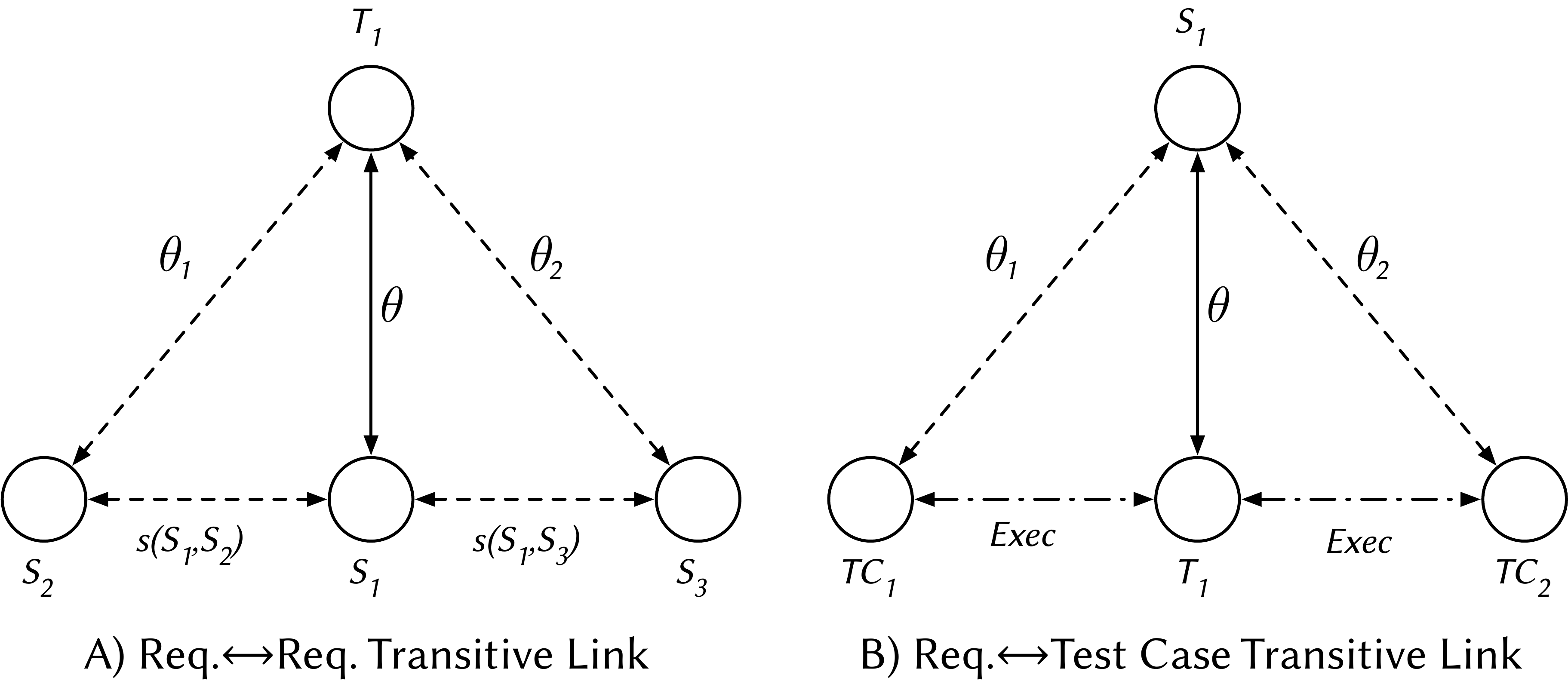}
\vspace{-0.7cm}
\caption{Illustration of Transitive Links}
\label{fig:trans-req2req}
\end{figure}

As discussed earlier, the probability that a trace link exists between a source and target artifact can be influenced by \textit{transitive} relationships among varying software development artifacts.  The third stage of \Comets HBN is able to utilize these transitive links to improve the accuracy of its inferred trace link. However, before we describe how our model reconciles this information in a probabilistic manner, it is first important to understand the phenomena of transitive links. At a high level, a transitive link is an inherent relationship between two software artifacts ($A_1$,$A_2$) that may influence the existence of a trace link between either $A_1$ and any other artifact or $A_2$ and any other artifact. \Comet is currently capable of leveraging two types of transitive links, one based on textual-similarities (req. $\leftrightarrow$ req.) and one based on dynamic execution information (req. $\leftrightarrow$ test case). However, \Comet could also be extended to model transitive relationships between other types of artifacts, such as commit messages or issues. Fig. \ref{fig:trans-req2req} provides an illustration of both transitive link types, which we detail below. Note that for execution traces, a relationship is considered \textit{strong} between a test method and source method if the test executes the method, and \textit{weak} otherwise. For req$\leftrightarrow$req relationships, it is \textit{strong} if the textual similarity is above the threshold $\tau$, and \textit{weak} if it is below the threshold.

\subsubsection{\textbf{Req. $\leftrightarrow$ Req. Links}} Consider $S_1,S_2,S_3$ as three source artifacts representing three discrete requirements documents and $T_1$ is a potential target document (\ie source code file), where the target relationship being inferred is $S_1\rightarrow T_1$, indicated by the solid line. The connections between nodes denote relationships among the artifacts. Consider the scenario in which the relationship between $S_1$ and $T_1$ is weak, but $S_1$ is highly similar to the other two requirements $S_2$ and $S_3$. Given that we know the three source artifacts are highly related, the relationships between $S_2\rightarrow T_1$ and $S_3\rightarrow T_1$ have a \textit{transitive} influence on the target relationship between $S_1$ and $T_1$. For instance, if $S_2\rightarrow T_1$ and $S_3\rightarrow T_1$ both indicate strong probabilities, then likewise the probability of the target link $S_1\rightarrow T_1$ should be increased to account for these transitive relationships.

\subsubsection{\textbf{Req.$\leftrightarrow$ Test Case Links}} Consider $S_1$ to be a source artifact representing a requirement document, $T_1$ to be a potential target source code file, and $TC_1,TC_2$ to be test cases, where the target relationship being inferred is $S_1\rightarrow T_1$, indicated by the solid line. Consider again the scenario in which the relationship between $S_1$ and $T_1$ is weak, whereas the relationship between $S_1$ and $TC_1,TC_2$ are stronger. If we observe that $TC_1$ and $TC_2$ are related to $T_1$ by execution information (\eg $TC_1$ and $TC_2$ both exercise $T_1$, \textit{Exec} in Fig. \ref{fig:trans-req2req}) then, this transitive relationship should influence the probability that a trace link exists between $S_1$ and $T_1$.

\subsubsection{\textbf{Incorporating Transitive Links into C{\footnotesize OMET}'s HBN}} In order for our HBN to incorporate transitive req. $\leftrightarrow$ req. links, it must first derive the set of requirements that are related to a given target requirement $S_x$. To accomplish this, one of several IR techniques can be used to compute textual similarity, or the first stage of our model can be used to derive the relationships, illustrated as $s(S_1,S_2)$ \& $s(S_1,S_3)$ in Fig. \ref{fig:trans-req2req}. To incorporate information from req. $\leftrightarrow$ test case links, dynamic information must be collected that provides the $Exec_1$ \& $Exec_2$ relationships illustrated in Fig. \ref{fig:trans-req2req}. In either case, a specified threshold $\tau$ signals whether a pair of requirements is related, and the total number of related requirements or test cases is specified by the hyper-parameter $\pi$.  Once the related requirements have been derived, our HBN estimates three hyperpriors, $w_t$, $B_{mix}$ and $\mu_{trans}$. First $w_t$ is formulated as a Dirichlet distribution according to the number of related transitive requirements. Then to estimate $B_{mix}$, the first stage of our HBN is computed between each related requirement and a given target artifact $T_y$. The inferred values for each transitive link, and $w_t$ are used to form a mixture model:

\vspace{-0.4cm}
\begin{equation}
	B_{mix} \sim Mix(w_t,\theta_\pi)
\end{equation}

\noindent where $B_{mix}$ is a Beta mixture model parameterized by the 1st stage inference of each transitive link and $\pi$ weights modeled as a Dirichlet distribution parameterized by $\pi$. This Dirichlet distribution is then used to derive a meditated normal distribution $\mu_{trans}$:

\vspace{-0.3cm}
\begin{equation}
	\mu_{trans} \sim \rho*B_{mix} + (1-\rho)*Mean[\theta_{IR}]
\end{equation}

\noindent where $Mean[\theta_{IR}]$ represents the mean of the probability distribution of IR similarity values (from stage 1) on the trace link prior and $\rho$ is represents the \textit{belief factor} of the transitive links (\eg the degree to which the transitive relationships should affect overall prior trace link probability). $\mu_{trans}$ can then be utilized to derive the reward and penalty functions introduced earlier where $\Upsilon = \sigma*(1-\mu_{trans})$ whereas $\Sigma = \sigma*\mu_{trans}$. The reward and penalty functions can then in turn be used to influence the mean of trace link prior $\mu$:

\vspace{-0.3cm}
\begin{equation}\label{eq:mean-affect-combined}
	\mu \sim N(\mu_n= \mu_{trans} + \Sigma + \Upsilon, sd=0.01)
\end{equation}

\noindent in the same manner as introduced in Eq. \ref{eq:mean-affect}. In this way, our model is capable of incorporating information from transitive links, increasing the overall prior probability if transitive links are strongly connected to the target artifact $T_y$ and decreasing it if they are not strongly connected.

\vspace{-0.2cm}
\subsection{Stage 4 - The Holistic Model}
\label{subsec:model-comp4}

The holistic model combines all three underlying stages. To accomplish this, the calculations of the reward and penalty functions for affecting the mean $\mu$ of the overall prior are modified to incorporate information from both expert feedback and transitive links:

\vspace{-0.3cm}
\begin{equation}
\begin{split}
	\Upsilon \sim (1-\mu_{trans})*\sigma*D \\
	\Sigma \sim \mu_{trans}*\sigma*(D-1)
\end{split}
\end{equation}

\noindent Then Eq. \ref{eq:mean-affect-combined} can be used to derive the new mean for the overall prior probability distribution of the model.

\subsection{Inferring the Posterior}
\label{subsec:model-posterior}

In order to reason about the probability that a trace link exists, we must estimate the posterior probability distribution of our hierarchical model $p(\Theta|L)$ according to the observable data $L$ and prior knowledge of the link $p(\Theta)$. Here $p(\Theta)$ encompasses the trace link prior and all constituent hyperpriors depending upon the stage of the model.  Once the posterior has been estimated, \Comet utilizes the \textit{mean} of the distribution as the general probability that a link exists. We can represent the general calculation of the posterior for our model using using Bayes Theorem as follows:  

\vspace{-0.3cm}
\begin{equation} \label{eq_bayes}
p(\Theta|L) = \dfrac{p(\Theta)p(L|\Theta)}{\int p(\Theta)p(L|\Theta)d\Theta} \propto p(\Theta) \prod\limits_{i=1}^n p(L_i|\Theta_i)
\end{equation}

\noindent where $n$ represents the total number of observations (\ie the number of underlying IR techniques and configurations).  \Comets HBN is non-trivial, and thus the posterior $p(\Theta|L)$ cannot be computed analytically. Therefore, we turn to approximation techniques for estimating the posterior probability distribution. Comet can currently utilize three different techniques including (i) Maximum a Posteriori (MAP) estimation~\cite{Bassett2018MaximumEstimators}, a Markov Chain Monte Carlo (MCMC) technique via the No-U-Turn sampling (NUTS) process~\cite{Hoffman2011TheCarlo}, and a machine learning-based technique called Variational Inference (VI)~\cite{Bishop:2006}. We provide experimental results in Sec. \ref{sec:results} for all techniques for Stage 1 of \Comets model, and NUTS/MAP for Stages 2-4, as VI cannot be applied to more complex stages of the model.

\subsection{\hspace{-0.1cm}
The C{\small OMET} Python Library \& Jenkins Plugin}
\label{subsec:comet-jenkins}

\begin{figure}[tb]
	\centering
	\vspace{-0.0cm}
	\includegraphics[width=\columnwidth]{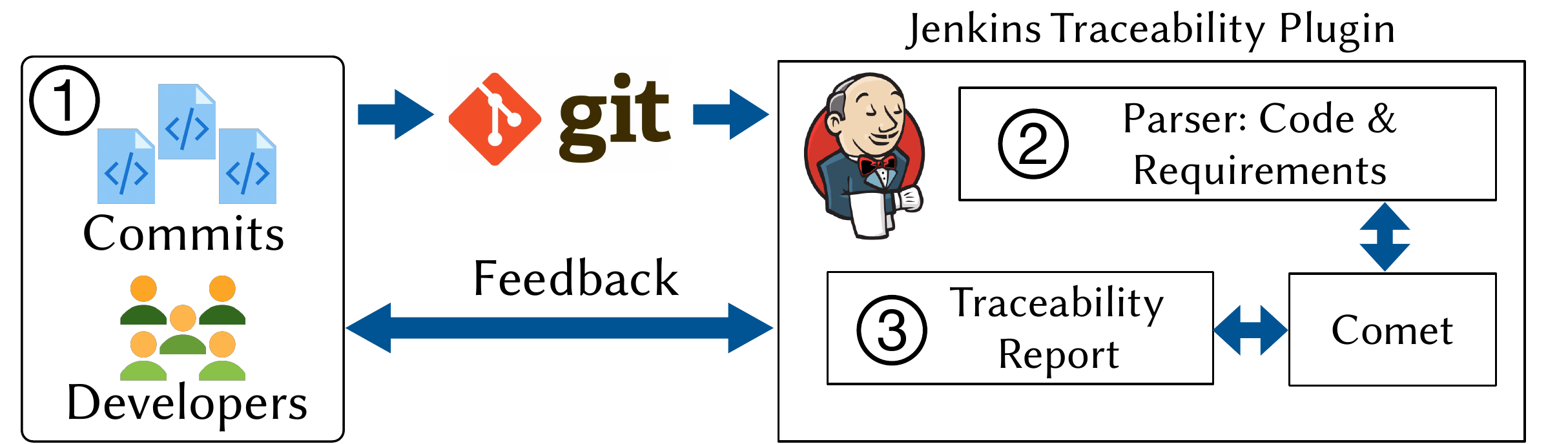}
	\vspace{-0.7cm}
	\caption{Comet Jenkins Plugin Flow}
	\label{fig:plugin-flow}
	\vspace{0.1cm}
\end{figure}

We implemented the four stages of \Comets HBN in an extensible traceability library written in \texttt{\textbf{\small python3}}. In order to explore the practical applicability of \Comet, we implemented the first two stages of \Comets HBN as a Jenkins~\cite{jenkins} plugin.  We did not implement the final two stages due to time limitations, but are actively working on this in partnership with Cisco. This plugin was developed by one of the authors during an internship with Cisco Systems in close collaboration with researchers, engineers and analysts. The code and extensive documentation for the \Comet library and plugin is available in our online appendix~\cite{appendix}.

\subsubsection{\textbf{An Illustrated Use Case of the C{\footnotesize OMET} Plugin}} \figref{fig:plugin-flow} provides a general overview of the \Comet plugin architecture. To illustrate the \textit{utility} of the plugin, we describe its workflow from the viewpoint of a developer. As depicted in \figref{fig:plugin-flow}-\textcircled{1} the plugin is triggered when developers commit changes to a given project configured to utilize our traceability plugin in Jenkins. The commit triggers a job that checks out, compiles, and executes the project code according to our industrial partner's existing CI pipeline. As illustrated in \figref{fig:plugin-flow}-\textcircled{2} our plugin parses and preprocesses the source code, test code, and requirement text to be analyzed with \Comet once the normal CI's build flow finishes. The preprocessed corpora of requirements, code, and tests are then passed to the \Comet library where the \textit{first stage} of the HBN (configured according to the tuning datasets described in Sec. \ref{sec:study}) is run to establish an initial set of trace links among artifacts.  Note that our plugin can be configured to run according to varying intervals (\eg every minor or major release). Given that our plugin supports the first two stages of \Comets HBN, it is capable of collecting feedback from developers to improve trace links. To do this, developers select an option from a dropdown likert scale, with each option representing a potential value of $\mu_d$ for stage 2 of \Comets HBN (Strongly agree=1.0, Agree=0.75, Unsure=0.5, Disagree=0.25, \& Strongly Disagree=0.0). To ensure a responsive feedback mechanism to reflect developer feedback, the Stage 2 link probabilities are precomputed. 

\begin{figure}[tb]
\centering
\vspace{-0.0cm}
\includegraphics[width=\columnwidth]{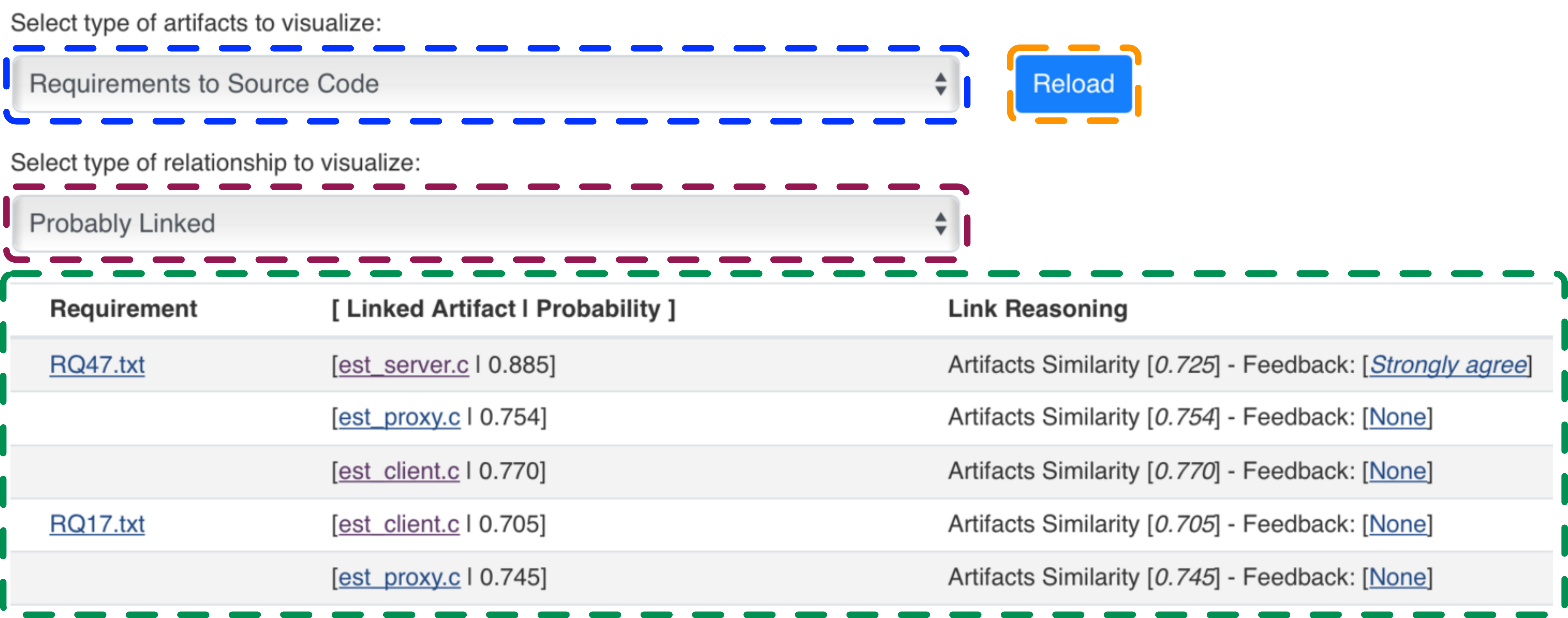}
\vspace{-0.7cm}
\caption{Comet Jenkins Plugin Traceability Report}
\label{fig:comet-plugin-ui}
\vspace{0.0cm}
\end{figure}

Once the plugin job finishes it generates an interactive \textit{traceability page} (Fig. \ref{fig:comet-plugin-ui}), that allows a developer or analyst to view inferred trace links and provide feedback. As shown in \figref{fig:comet-plugin-ui} the blue rectangle highlights a drop-down menu that allows a developer to filter inferred trace links by type (\ie req$\leftrightarrow$src, req$\leftrightarrow$test cases, src$\leftrightarrow$test cases, and artifacts not linked). The drop-down menu highlighted in red allows developers to filter trace links according to the inferred probability ranges of $\theta$ in \Comets HBN including: \textit{probably linked} (\ie  $\theta=[0.7,1)$), \textit{probably not linked} (\ie $\theta=[0,0.4)$), and \textit{unsure} representing links where the model is unable to make a confident inference ( \ie $\theta=[0.4,0.7)$). The area highlighted in green lists the types of selected source and target artifacts allowing the user to easily inspect candidate trace links as well as provide feedback on the inferences. When a developer clicks on the \texttt{\small feedback} link for a given pair of artifacts, a popup window allows them to select one of the likert options delineated earlier. Finally, the traceability page also allows developers to view artifacts that are not linked to any other artifacts, and thus may be suspicious. Additional screenshots detailing the workflow of the \Comet plugin are available in our online appendix~\cite{appendix}.

\subsubsection{\textbf{C{\footnotesize OMET}'s Complexity \& Scalability}}

We have designed the \Comet plugin and library to facilitate easy integration into modern automated CI/CD systems via a set of higher level APIs that abstracts much of the complexity of the model for most users, while still providing mechanisms for advanced users to tweak parameters. Furthermore, \Comet was able to be successfully deployed into a commercial CI pipeline for a pilot project with our industrial partner. We have also designed the \Comet plugin and library to be highly scalable. Trace link probabilities between each pair of artifacts can be calculated independently, making the process highly parallelizable. Thus, we implemented parallel process management using \texttt{Theano}~\cite{theano} into the \Comet library, allowing computation to scale across modern multi-core machines. To further optimize performance, the \Comet plugin can makex use of change analysis in git to only recompute trace link values for artifacts that have been altered since the last computation.

%\vspace{-0.2cm}
%--------------------
\section{Design of the Experiments}
\label{sec:study}

	To evaluate \Comet, we perform an extensive empirical evaluation with two major \textit{goals}: (i) evaluate the effectiveness of the four stages of \Comets HBN in terms of their ability to effectively infer trace links, and (ii) examine whether \Comet is applicable in industrial workflows. The \textit{quality focus} of our study is \Comets effectiveness, in terms of generating an accurate and complete set of trace links, and practical applicability. We formulate the following set of RQs:

\begin{itemize}

	\item{\textbf{RQ$_1$}: \textit{How effective is \Comet at inferring candidate trace links using combined information from IR/ML techniques?}}

	\item{\textbf{RQ$_2$}: \textit{To what extent does expert feedback impact the accuracy of the candidate trace links of \Comet?}}

	\item{\textbf{RQ$_3$}: \textit{To what extent does information from transitive links improve \Comets trace link inference accuracy?}}

	\item{\textbf{RQ$_4$}: \textit{How effective is the holistic \Comet model in terms of inferring candidate trace links?}}

	\item{\textbf{RQ$_5$}: \textit{Do professional developers and security analysts find our implementation of the \Comet Jenkins plugin useful?}}
	
\end{itemize}

%--------------------
\subsection{Experimental Context}
\label{subsec:exp-context}

%--------------------
\subsubsection{\textbf{Subject Datasets}}
\label{subsec:exp-datasets}

\begin{table}[tb]
	\footnotesize
	\centering
	%	\small
	\vspace{-0.0cm}
	\caption{Datasets used for \Comets evaluation. Rq = Requirement, Src = Source code, UC = Use Case}
	\vspace{-1.0em}
	\label{tab:datasets}
	\setlength{\tabcolsep}{0.1em}
\begin{tabular}{|C{1.45cm}| C{1.1cm} |C{1.2cm}| C{1.0cm}| C{1.0cm}| C{1.2cm}| C{1.1cm}|}
\hline
\rowcolor[HTML]{C0C0C0} 
\textbf{Project}  & \textbf{Language}    & \textbf{Size (LoC)} & \textbf{\# Source Artifacts} & \textbf{\# Target Artifacts} & \textbf{\#Pairs/ \#Links}  & \textbf{Artifact Type} \\ \hline
\multicolumn{7}{|c|}{\cellcolor[HTML]{C0C0C0}\textbf{Tuning Projects}}                                                                                                   \\ \hline
Albergate          & Java                 & 10,464      &  55  &  17  &     935 / 53       & Rq$\rightarrow$Src          \\ \hline
EBT                  & Java                 & 1,747     &  40  &  25  &     1000 / 51     & Rq$\rightarrow$Tests          \\ \hline\hline
\rowcolor[HTML]{C0C0C0}
\multicolumn{7}{|c|}{\textbf{Experimental Projects}}                                                                                              \\ \hline
\multirow{2}{*}{LibEST} & \multirow{2}{*}{C}   & \multirow{2}{*}{70,977} &  59  &  11  &  649 / 204  & Rq$\rightarrow$Src            \\ \cline{4-7} 
                        & &                                      &  59       &            18          &     1,062 / 352                    & Rq$\rightarrow$Tests          \\ \hline
eTour      & Java                 & 23,065     &  58  & 116 &     6,728 / 308        & UC$\rightarrow$Src             \\ \hline
EBT        & Java                 & 1,747      &  40  & 50 &      2,000 / 98       & Rq$\rightarrow$Src            \\ \hline
SMOS       & Java                 & 9,019      &  67  & 100 &     6,700 / 1,044        & UC$\rightarrow$Src             \\ \hline
iTrust     & Java, JSP 			  & 38,087     &  131 & 367 &     48,077 / 399        & Rq$\rightarrow$Src            \\ \hline
\end{tabular}
\vspace{0.1cm}
\end{table}

The \textit{context} of this empirical study includes the eight datasets shown in \tabref{tab:datasets}. Six of these are taken from the open source CoEST community datasets~\cite{coest-datasets}. These datasets represent a set of benchmarks created by the research community and widely used as an effective assessment tool for automated traceability techniques~\cite{Antoniol:e,Cleland-Huang:TSE'03,Poshyvanyk:TEFSE'11,Gethers:ICSM'11}. In order to maintain the quality of our experimental subjects, we do not use all available projects in the CoEST repository, as we limited our studied systems to those that: (i) included trace links from requirements or use cases written in natural language to some form of code artifact, (ii) were written in English and/or included English translations, and (iii) had at least 1k LoC.  We utilize two datasets to investigate and tune the hyper-parameters of \Comets HBN, Albergate, and the Rq$\rightarrow$Tests dataset of the EBT project. We utilize the other six datasets for our empirical evaluation. The subject system called ``LibEST'' is an open source networking related software project, which was created and is actively maintained by engineers at Cisco as an implementation of RFC-7030 ``Enrollment over Secure Transport''. We derived the ground truth set of trace links between Rq$\rightarrow$Src and Rq$\rightarrow$Tests for this dataset in close collaboration with our industrial partner. First, one of the authors carefully created an initial set of trace links. Then, an engineer working on the project reviewed the links and confirmed or denied a subset, based on their availability. The author then revised the links using the engineer's feedback, and this process continued over several months until the ground truth was established. The "LibEST" dataset is available along with all of our experimental data to facilitate reproducibility~\cite{appendix}.

\vspace{-0.0cm}
%--------------------
\subsubsection{\textbf{Studied IR Techniques}}
\label{subsec:exp-datasets}

The ``base'' first stage of \Comets HBN is able to utilize and unify information regarding the textual similarity of development artifacts as computed by a set of IR/ML techniques.  While there is technically no limit to the number of IR/ML techniques that can be utilized, we parameterized our experiments using ten IR-techniques enumerated in Table~\ref{tab:ir-techniques}. The first five techniques are standalone techniques, whereas the second five are combined techniques utilizing the methodology introduced by Gethers \etal~\cite{Gethers:ICSM'11}. 
This combined approach normalizes the similarity measures of two IR techniques and combines the similarity measures using a weighted sum. We set the weighting factor $\lambda$ for each technique equal to 0.5, as this was the best performing configuration reported in the prior work~\cite{Gethers:ICSM'11}. We explain the differences between the technique employed by Gethers et. al. and \Comet in Sec. \ref{sec:related-work}. The other parameters for each of the techniques were derived by performing a series of experiments on the two tuning datasets, and using the optimal values from these experiments. For all IR techniques, we preprocessed the text by removing non-alphabetic characters and stop words, stemming, and splitting camelCase. Text was vectorized using tf-idf vectors. We explored using word2vec to encode documents but found the performance to be considerably worse than tf-idf vectors. We performed 30 trials for each technique involving LDA, and chose the number of topics that led to optimal performance on our tuning projects. To aid in experimental reproducibility, complete configurations for each technique are listed in our online appendix~\cite{appendix}.

\begin{table}[tb]
	\footnotesize
	\centering
	%	\small
	\vspace{-0.0cm}
	\caption{IR/ML Techniques used in the Construction of C{\tiny OMET}'s HBN}
	\vspace{-1.0em}
	\label{tab:ir-techniques}
	\setlength{\tabcolsep}{0.1em}
\begin{tabular}{|C{4.2cm}| C{1.6cm} |C{2.0cm}|}
\hline
\rowcolor[HTML]{C0C0C0} 
\multicolumn{1}{|c|}{\cellcolor[HTML]{C0C0C0}\textbf{IR Technique}} & \multicolumn{1}{c|}{\cellcolor[HTML]{C0C0C0}\textbf{Tag}} & \multicolumn{1}{c|}{\cellcolor[HTML]{C0C0C0}\textbf{Treshold Technique}}                        \\ \hline
Vector Space Model                                                  & VSM                                                       & Link-Est                                                                                                 \\ \hline
Latent Semantic Indexing                                            & LSI                                                       & Link-Est                                                                                                  \\ \hline
Jensen-Shannon Divergence                                  & JS                                                        & Min-Max                                                                                            \\ \hline
Latent Dirichlet Allocation                                         & LDA                                                       & Min-Max                                                                            \\ \hline
NonNegative Matrix Factorization                                    & NMF                                                       & Median                                                                                          \\ \hline
Combined VSM + LDA                                                  & VSM+LDA                                                   & Link-Est                                                       \\ \hline
Combined JS+LDA                                                     & JS+LDA                                                    & Link-Est                                                 \\ \hline
Combined VSM+NMF                                                    & VSM+NMF                                                   & Link-Est                                                                    \\ \hline
Combined JS+NMF                                                     & JS+NMF                                                    & Link-Est                                                            \\ \hline
Combined VSM+JS                                                     & VSM+JS                                                    & Min-Max                                                                             \\ \hline
\end{tabular}
\end{table}

%--------------------
\vspace{-0.0cm}
\subsection{RQ$_1$: C{\footnotesize OMET} Performance w/ Combined IR/ML Techniques}
\label{subsec:study-rq1}

To answer RQ$_1$, we ran the first stage of \Comets HBN on our six evaluation datasets using the ten IR/ML techniques enumerated in Table \ref{tab:ir-techniques}. However, as explained in Section~\ref{subsec:model-comp1}, in order to accurately estimate the likelihood function $Y$ we need to choose a threshold $k_i$ for each IR technique that maximizes the precision and recall of the trace links according to the computed textual similarity values. To derive the best method for determining the threshold for each IR technique, we performed a meta evaluation on our two tuning datasets. We examined five different threshold estimation techniques: (i) using the mean of all similarity measures for a given dataset, (ii) using the median of all similarity measures across a given dataset, (iii) using a Min-Max estimation, (iv) a sigmoid estimation, and (v) link estimation (Link-Est), where an estimation of the number of confirmed links for a dataset is made based on the number of artifacts, and a threshold derived to ensure that the estimated number of links is above that threshold.  We performed each of these threshold estimation techniques for all studied IR techniques across our two tuning datasets, and compared each estimation to the known optimal threshold. We used the optimal technique across our two tuning datasets, as reported in Table \ref{tab:ir-techniques}. To aid in reproducibility, we provide a detailed account of these experiments in our online appendix~\cite{appendix}. 

To provide a comparative baseline against which we can measure \Comets performance, we report results for the best-performing and median of the studied IR/ML techniques, optimally configured for each dataset. We chose to optimally configure the baseline techniques, even though such configurations would not be possible in practice due to the absence of a ground truth, in order to illustrate how close Comet can come to the ``best-case baseline scenario''. 

To provide a comprehensive comparison of \Comet to a state of the art technique for candidate trace link generation, we re-implemented the DL-based approach proposed by Guo \etal\cite{Guo:ICSE'17}. However, it should be noted that the intended purpose of this DL approach and Comet differ. The DL technique proposed by Guo \etal was intended to be both trained and evaluated on a single project that contains a set of \textit{pre-existing} trace links the model can be trained upon, and was quite effective in improving the accuracy of trace links in this scenario. However, as pre-existing trace links may not always exist \Comet \textit{does not} require them for analysis. Instead, our experiments aim to illustrate that \Comet can accurately infer trace links when tuned on one small set of projects, and applied to others. Therefore, we design an experimental setup where both techniques are applied on projects without pre-existing trace links. Thus, we train the DL approach on our two tuning projects, using the optimal parameters reported in~\cite{Guo:ICSE'17}. Our main goal in comparing with this DL technique is to illustrate the performance of a recent ML-based technique applied to Comet's intended ``cold-start'' use case.

In order to measure the performance of our studied techniques for inferring trace links, we utilize three main metrics, Precision, Recall, and Average Precision (AP), similar to prior work that evaluates automated traceability techniques \cite{Gethers:ICSM'11,Guo:ICSE'17}. Given that candidate link generation techniques infer a probability or similarity that a trace link exists, a threshold similarity or probability value must be chosen to make the final inference. In order to summarize the performance of our studied techniques, we calculate the Average Precision as a weighted mean of precisions per threshold: $AP = \Sigma_n(R_n-R_{n-1})P_n$ where $P_n$ and $R_n$ are the Precision and Recall at the $n$th threshold. Thus, the AP provides a metric by which we can quantitatively compare the performance of different approaches. For the results of \Comet, we report the highest AP achieved by the posterior estimation techniques outlined in Sec. \ref{subsec:model-posterior}. In addition to AP, we also provide Precision/Recall (P/R) curves to illustrate the trade-off between precision and recall at different threshold values. Curves further away from the origin of the graph indicate better performance. In lieu of a non-parametric statistical test as suggested by recent work~\cite{Furia:TSE'19}, we perform a confidence interval analysis~\cite{Neyman:37} between our baseline techniques and Stage 1 of \Comet by calculating the standard error across different threshold values, applying bootstrapping where necessary. Thus, if one technique outperforms another within the bounds of our calculated error, it serves as a strong indication of statistical significance.

%--------------------
\vspace{-0.0cm}
\subsection{RQ$_2$: C{\footnotesize OMET} Performance w/Expert Feedback}
\label{subsec:study-rq2}

Collecting \textit{actual} developer feedback on trace links for each of our test datasets was not possible  given the time constraints on developers from our industrial partner, and we did not have access to the developers of the other projects. Thus, in order to evaluate Stage 2 of \Comets HBN, we simulated developer feedback by randomly sampling 10\% of the artifact pairs from each studied subject, and used the ground truth to provide a confidence level for each of the sampled links.  To accomplish this, we provided the model with a confidence value $c$ of 0.9 if a link existed in the ground truth, and 0.1, if the link did not exist. However, even trace links derived from experts can be error-prone. Hence, we performed three types of experiments to simulate imperfect links being suggested to our model. That is, for the set of randomly sampled links, we intentionally reversed the confidence values according to the ground truth, for 25\% and 50\% of the sampled links respectively to simulate varying degrees of human error in providing link feedback. In other words, we sampled a small number of trace links from the ground truth, and then used these links to confirm/deny links predicted by \Comet (i.e., if a ground truth link existed, and \Comet predicted it, then it was confirmed). Because developers may not be correct all of the time, we simulated this by randomly flipping the sampled ground truth, which has a similar effect to a developer incorrectly classifying certain predicted links.

We set the value for the \textit{belief factor} of the developer feedback $\sigma=0.5$. For these experiments we illustrate the impact of developer feedback on AP and P/R curves for \textit{only} the sampled links. In addition to the baseline IR techniques described in the procedure for RQ$_1$, we also compare our results from Stage 2 of the model to Stage 1, to illustrate the relative improvement.

%--------------------
\vspace{-0.0cm}
\subsection{RQ$_3$: C{\footnotesize OMET} Performance w/Transitive Links}
\label{subsec:study-rq3}

To measure the impact that transitive links have on the trace link inference performance of Stage 3 of \Comets HBN, we examined the impact of transitive links between requirements as described in Sec. \ref{subsec:model-comp3}. We utilize transitive requirement links rather than transitive links established by execution traces, as only one of our datasets (LibEST) had executable test cases. To derive the transitive relationships between artifacts, we computed the VSM similarity among all source documents for each dataset (\eg requirements, use cases) and explored two values for the threshold $\tau$, 0.65, and 0.5. We derived these thresholds by examining the total number of transitively linked requirements in our tuning datasets to achieve a balance between too many and too few requirements being linked. We set the \textit{belief factor} $\rho$ for Stage 3 of the HBN equal to 0.5. We report results for these experiments for only those requirements where transitive links impacted \Comets performance.

%--------------------
\vspace{-0.0cm}
\subsection{RQ$_4$: Holistic C{\footnotesize OMET} Performance}
\label{subsec:study-rq4}

To evaluate the overall performance of \Comets holistic model, we combined our experimental settings for RQ$_2$ \& RQ$_3$. That is, we randomly sampled 10\% of the links from each dataset and simulated developer feedback with a 25\% error rate. Additionally, we incorporated transitive links between requirements using the same procedure outlined for RQ$_3$. For the transitive links, we set $\tau$ to 0.65, and we set the $\sigma$ and $\rho$ hyper-parameters both equal to 0.5. For this research question, we report results across all links. 

%--------------------
\vspace{-0.0cm}
\subsection{RQ$_5$: C{\footnotesize OMET} Industrial Case Study}
\label{subsec:study-rq6}

Given that the ultimate goal of designing \Comet is for the approach to automate trace link recovery within industry, we perform a case study with our industrial partner. This case study consisted of two major parts.
First, we conducted a feedback session with six experienced developers who have been contributing to the LibEST subject program. This session consisted of a roughly 15 minute presentation introducing the \Comet Jenkins plugin. Then the developers were asked to use the plugin, which had been configured for LibEST, and evaluate the links and non links for which the model was most confident (\ie the highest and lowest inferred probabilities). Then after using the tool, they were asked a set of likert-based user experience (UX) questions derived from the SUS usability scale by Brooke \cite{Brooke:96}. Additionally, participants were asked free-response user preference questions based on the honeycomb originally introduced by Morville~\cite{Morville:04}. 
Second, we conducted semi-structured interviews with two groups consisting of roughly 15 engineering managers who specialize in auditing software for security assurance. During these interviews, a video illustrating the \Comet plugin was shown, and a discussion was conducted with the questions illustrated in Fig. \ref{fig:LibEST-study}. We report results from both studies.

%--------------------
%\vspace{-0.2cm}
\section{Empirical Results}
\label{sec:results}

This section presents the results for our five proposed RQs. We highlight two P/R curves and focus our discussion on the AP results. However, all P/R curves and confidence interval graphs are currently available in our appendix alongside all experimental data~\cite{appendix}.

%--------------------
\vspace{-0.0cm}
\subsection{RQ$_1$ Results: C{\footnotesize OMET} Stage 1 Performance}
\label{subsec:results-rq1}

\begin{table}[tb]
	\footnotesize
	\centering
	%	\small
	\vspace{-0.0cm}
	\caption{\footnotesize AP Results from Stages 1 \& 4 of \Comet. The given $p$ values from the Wilcoxon test measure the significance of performance variations between Stage 1, and Stage 4 of \Comets model compared to the median (Med.) baseline of IR/ML techniques. ``I=Net'' signifies the ``Industry-Net'' dataset.}
	\vspace{-1.0em}
	\label{tab:stage1-4-results}
	\setlength{\tabcolsep}{0.1em}
\begin{tabular}{|c?c?c|c?c?c|c?c|}
\hline
\rowcolor[HTML]{C0C0C0} 
\textbf{Subject}                  & \textbf{Best Base.} & \textbf{Med. Base.} & \textbf{Std. Err} & \textbf{~DL~} & \textbf{Stage 1} & \textbf{Std. Err} & \textbf{Stage 4} \\ \hline
LibEST (Rq$\rightarrow$Src)        & 0.69  & 0.55 & $\pm$0.008   &  0.28   & 0.63  &  $\pm$0.006  & 0.64    \\ \hline
LibEST (Rq$\rightarrow$Tests)      & 0.42    & 0.36 &  $\pm$0.001   & 0.32  & 0.38   &  $\pm$0.002   & 0.42          \\ \hline
eTour                             & 0.40     & 0.30  &  $\pm$0.011   &   0.05   & 0.38    & $\pm$ 0.002 & 0.36            \\ \hline
EBT                               & 0.17  & 0.14 &   $\pm$0.005  & 0.07   & 0.15   &   $\pm$0.001   & 0.17           \\ \hline
SMOS                              & 0.29  & 0.25  &  $\pm$0.003  &  0.16 & 0.25  &  $\pm$ 0.001     &   0.27              \\ \hline
iTrust                            & 0.17    & 0.13  &  $\pm$0.006   &   0.01  & 0.17  &  $\pm$0  &  0.17    \\ \hline
\end{tabular}
\vspace{-0.0cm}
\end{table}

The AP values for Stage 1 of \Comets HBN are provided in Table \ref{tab:stage1-4-results} alongside the $p$ values for the Wilcoxon test between Comet and the median IR/ML baseline. The P/R curves for the iTrust dataset are illustrated in \figref{fig:pr1-results}. As Table \ref{tab:stage1-4-results} indicates, Stage 1 of \Comet outperforms the median IR/ML baseline across all subjects, to a statistically significant degree according to the confidence intervals. In some cases, such as for iTrust, LibEST, and eTour, Stage 1 of \Comet \textit{significantly} outperforms the median IR/ML baseline, and approaches the performance of the \textit{best} IR/ML baseline. \figref{fig:pr1-results} illustrates the P/R curve for the iTrust project, with performance that outpaces the best IR/ML technique, particularly for lower recall values. \Comet also outperforms the state of the art DL approach across all subjects, likely because the DL approach had difficulty generalizing semantic relationships across datasets.

These results signal remarkably strong performance for \Comets Stage 1 model. Recall that, the Stage 1 model \textit{only} utilizes observations taken from the set of ten IR/ML techniques introduced in Sec. \ref{subsec:study-rq1}, thus the fact that the Stage 1 model was able to consistently outperform the median IR/ML baselines and in some cases, nearly match the best IR/ML baseline. This indicates that \Comets HBN is capable of effectively combining the observations from the underlying IR/ML techniques for improved inference power. This is significant, as currently practitioners cannot know a-priori which IR/ML technique for traceability will perform best on a given project without pre-existing trace links. Thus, by combining the collective information of several IR techniques \Comets first stage HBN is able to perform \textit{\textbf{consistently well}}, achieving reasonably high performance \textit{\textbf{across projects}}, lending to the credibility of using Comet for projects that do not contain preexisting links. 

%--------------------
\vspace{-0.0cm}
\subsection{RQ$_2$ Results: C{\footnotesize OMET} Stage 2 Performance}
\label{subsec:results-rq2}

\begin{table}[tb]
	\footnotesize
	\centering
	%	\small
	\vspace{-0.0cm}
	\caption{\footnotesize AP Results from Stage 2 of \Comet with simulated expert feedback with error rates of 25\% and 50\%. The Baseline AP reported in this table is the median of the IR/ML techniques for the sampled links affected by feedback.}
	\vspace{-1.0em}
	\label{tab:stage2-results}
	\setlength{\tabcolsep}{0.1em}
\begin{tabular}{|c?c|c|c?c|c|c|}
\hline
\rowcolor[HTML]{C0C0C0} 
\textbf{Subject}           & \textbf{Baseline} & \textbf{St.1} & \textbf{St.2 (25\%E)}  & \textbf{Baseline} & \textbf{St.1} & \textbf{St.2 (50\%E)}  \\ \hline
LibEST (Rq$\rightarrow$Src)   & 0.52   & 0.65  & 0.96             & 0.52   & 0.65  & 0.64      \\ \hline
LibEST (Rq$\rightarrow$Tests) & 0.28   & 0.32 & 0.80          & 0.28   & 0.32  & 0.44    \\ \hline
eTour                      & 0.48  & 0.60  & 0.66           & 0.48 & 0.60 & 0.39     \\ \hline
EBT                        & 0.20  & 0.22   & 0.38                      & 0.20   & 0.22   & 0.24     \\ \hline
SMOS                       & 0.18    &  0.17  &  0.39         & 0.18   & 0.17   & 0.17              \\ \hline
iTrust                     & 0.12  &  0.15 & 0.25                       & 0.12   &  0.15  &  0.10            \\ \hline
\end{tabular}
\vspace{0.35cm}
\end{table}
\begin{table}[tb]
	\footnotesize
	\centering
	%	\small
	\vspace{-0.7cm}
	\caption{\footnotesize AP Results from Stage 3 of \Comet for transitive links between requirements with $\tau$=0.55 and $\tau$=0.65. The baseline reported in this table is the median of the IR/ML techniques for links affected by transitive relationships.}
	\vspace{-1.0em}
	\label{tab:stage3-results}
	\setlength{\tabcolsep}{0.1em}
\begin{tabular}{|c?c|c|c?c|c|c|}
\hline
\rowcolor[HTML]{C0C0C0} 
\textbf{Subject}           & \textbf{Baseline} & \textbf{St. 1} & \textbf{St.3 ($\tau$=.55)} & \textbf{Baseline} & \textbf{St. 1} & \textbf{St.3 ($\tau$=.65)} \\ \hline
LibEST (Rq$\rightarrow$Src)   & 0.53   & 0.6    & 0.59           & 0.39    & 0.67   & 0.44               \\ \hline
LibEST (Rq$\rightarrow$Tests) & 0.38   & 0.4  & 0.38             & 0.18  & 0.19  & 0.22               \\ \hline
eTour                      & 0.33  & 0.4 & 0.42                & 0.37  & 0.48 & 0.48                \\ \hline
EBT                        & 0.24  & 0.26   & 0.24                        & 0.02  & 0.03   & 0.06                \\ \hline
SMOS                       &  0.19  &  0.20   &  0.19                              & 0.24    & 0.23  & 0.24                \\ \hline
iTrust                     &  0.11   &  0.14  &  0.15                                &  $-$  &  $-$    &  $-$          \\ \hline
\end{tabular}
\end{table}

The AP for for Stage 2 of \Comet across all subject programs for both 25\% and 50\% error rates is given in Table \ref{tab:stage2-results}.The results indicate that Stage 2 of \Comets HBN is able to effectively incorporate expert feedback to improve the accuracy of its trace link inferences, as the Stage 2 model dramatically outperforms the median (and best) IR/ML techniques as well as the first stage of the model, with a simulated error rate of 25\%.  Even for the larger error rate of 50\%, we see Stage 2 outperform Stage 1 for LibEST (Rq$\rightarrow$Src), LibEST (Rq$\rightarrow$Tests) and EBT, while it slightly underperforms the Stage 1 model for the other subjects. These results illustrate that Stage 2 of \Comets HBN is able to effectively utilize expert feedback to improve its inferences, even in the presence of significant noise. 

%--------------------
\vspace{-0.0cm}
\subsection{RQ$_3$ Results: C{\footnotesize OMET} Stage 3 Performance}
\label{subsec:results-rq3}

\begin{figure}[tb]
\centering
\vspace{-0.0cm}
\subfloat[\footnotesize P/R Curve for iTrust for Stage 1.]{
\includegraphics[clip,width=0.9\columnwidth]{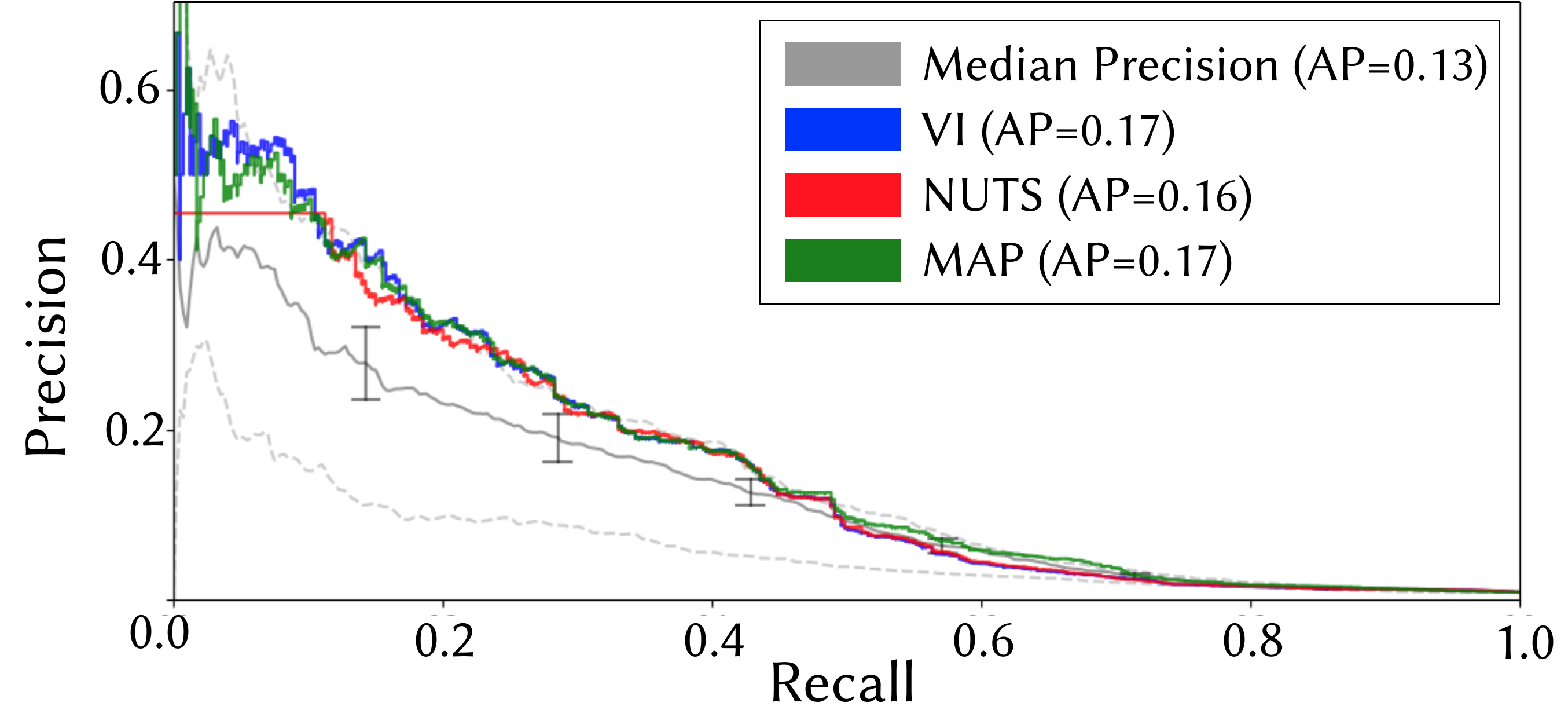}
\vspace{-0.85cm}
\label{fig:pr1-results}}
\vspace{-0.3cm}
\subfloat[\footnotesize P/R Curve for I-Net (Req$\rightarrow$Src) for Stage 4]{
\vspace{-0.35cm}
\includegraphics[clip,width=0.9\columnwidth]{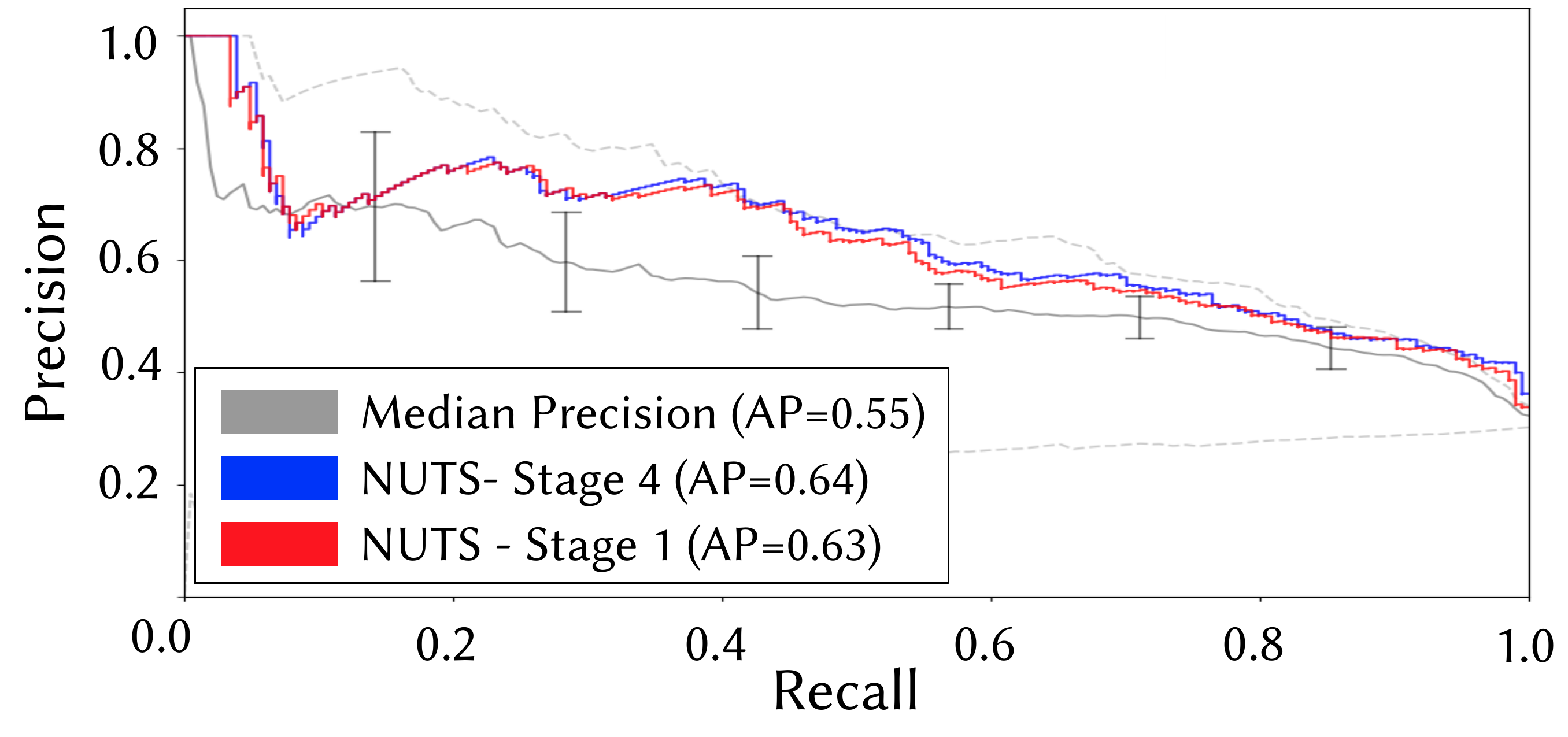}
\vspace{-0.45cm}
\label{fig:pr4-results}}
\vspace{-0.0cm}
\caption{\footnotesize Selected P/R Curves for Stage 1 and Stage 4 of \Comet.  Solid grey line is median of the baseline IR/ML techniques, dotted grey lines are best and worst performing IR/ML techniques respectively.}
\vspace{0.3cm}
\end{figure}

The AP results for Stage 3 of \Comet, which incorporates transitive relationships between requirements, for both $\tau=0.55$ and $\tau=0.65$ are given in Table \ref{tab:stage3-results} (There were no transitive links in iTrust for $\tau=0.65$). This table also includes the median of the baseline IR/ML techniques, as well as the Comet Stage 1 model AP results, for the set of links affected by transitive relationships (hence the differing Stage 1 columns). The results show that, in general, for $\tau=0.65$ for \Comets Stage 3 model, the accuracy of \Comets inferred trace links improve, with four of the six datasets showing improvements. For $\tau=0.55$ the results generally exhibit similar or slightly worse performance compared to Stage 1.  The fact that the higher value of $\tau$ led to better performance improvements is not surprising, as this parameter essentially controls the \textit{degree of relatedness} required to consider transitive relationships. Thus, a higher value of $\tau$ means that only highly similar transitive requirement relationships are considered by \Comet's model. Using a lower value for this parameter might introduce noise by incorporating transitive relationships between artifacts that don't have as high a degree of similarity. 

The LibEST (Rq$\rightarrow$Src) dataset exhibited decreased performance for $\tau=0.65$, however this is likely because the requirements for this industrial dataset are based on formal format from the Internet Engineering Task Force (IETF). The somewhat repetitive nature of the language used in these requirements could lead to non-related requirements being transitively linked, leading to a decrease in performance. This suggests leveraging transitive relationships between requirements leads to larger performance gains for more unique language. Overall, our results indicate that \Comets Stage 3 model improves the accuracy of links for a majority of subjects.

%--------------------
\vspace{-0.0cm}
\subsection{RQ$_4$ Results: C{\footnotesize OMET} Holistic Performance}
\label{subsec:results-rq4}

The AP results for the the holistic \Comet (Stage 4) model are given in Table \ref{tab:stage1-4-results}. These results show that \Comets holistic model outperforms the baseline median IR/ML techniques, and Stage 1 for all subject programs. For three subjects (LibEST Req$\rightarrow$Src, EBT, and iTrust), Comet's holistic model matches or outperforms the best baseline IR/ML technique. \figref{fig:pr4-results} illustrates the P/R curve for the LibEST (Req$\rightarrow$Src) dataset, which shows that the performance gains in inference precision extend for a large range of recall values. The results of these experiments demonstrate that \Comets holistic model is able to effectively combine information from multiple sources to improve its trace link inference accuracy.

%--------------------
\vspace{-0.2cm}
\subsection{RQ$_5$ Results: Industrial Case Study}
\label{subsec:results-rq6}

\begin{figure}[tb]
\centering
\vspace{0.2cm}
\includegraphics[width=\columnwidth]{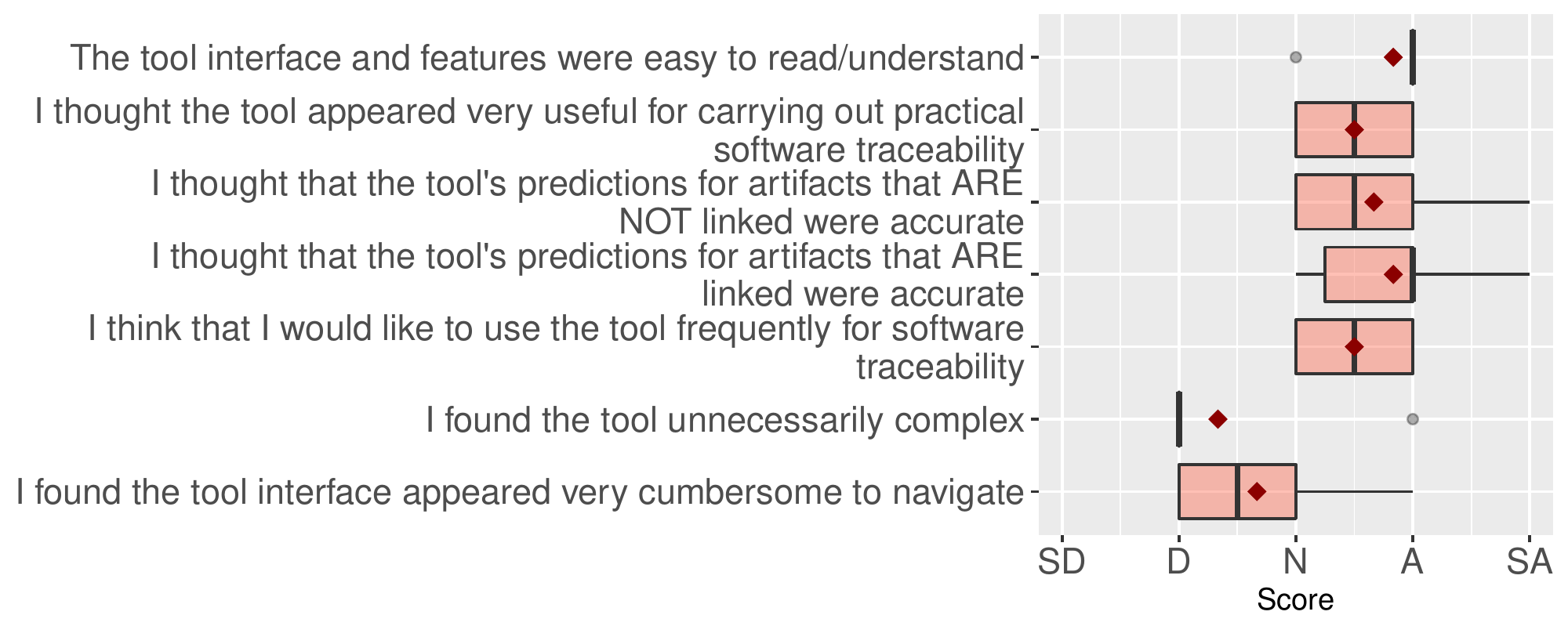}
\vspace{-0.7cm}
\caption{Results for LibEST Case Study UX Questions}
\vspace{-0.0cm}
\label{fig:LibEST-study}
\end{figure}

\figref{fig:LibEST-study} provides the responses to the likert-based UX questions from the six developers who work on the LibEST project after interacting with the \Comet plugin. Overall, the responses from these developers were quite positive. They generally agreed the \Comet plugin easy to use and understand, but more importantly, generally found the accuracy of the inferred links and non-links to be accurate. Additionally, we highlight representative responses to the user experience questions in this section, and provide the survey questions with response summaries in our online appendix~\cite{appendix}, in accordance with the NDA established with our industrial partner. Overall the developer responses were encouraging, indicating the practical need for approaches like \Comet. For instance, one developer stated their need for such a tool, \textit{``I really want a tool that could look at test cases and requirements and tell me the coverage. That way the team can know whether we are missing functionality or not.''} Another developer explained the need for a feature that incorporates developer feedback, stating the importance of the \textit{``ability to describe or explain how the code matches up with the code for future reference. Discussion/comments about such explanation as different developers might see links that others don't''}, whereas another developer stated, \textit{``Being able to provide feedback is useful and seeing it update the percentage immediately was nice.''}  This indicates that the support for developer feedback and responsiveness of the \Comet plugin inherently useful. Developers also found the traceability report to be useful, with most criticism suggesting practical UI improvements. For instance, developers appreciated \textit{``The fact that there were the three different options for viewing the traceability between different [artifacts]''}, and \textit{``The ability to bring up the specific requirement quickly in the same window.''}. These responses illustrate the utility that developers saw in the \Comet plugin. Given that these developers had little automated support for traceability tasks, they appreciated any automated assistance. 

We also collected feedback that validated the importance of the practical use cases that the \Comet plugin enabled. In these interviews, the teams generally stated that \Comet would be very useful for code auditing, as one manager stated that it would \textit{``allow compliance analysts to [inspect] links, look at the code and validate [the links]''}. Furthermore, a team responsible for security audits of systems found an interesting use case for \Comet that is often overlooked in traceability analysis. That is, they were interested in code and requirements that are \textit{not linked to any other artifact}, as such artifacts are likely to be suspicious and should be inspected further. In this case, \Comets inferences of non-links would be just as important as the inferences of links. Overall, the interviewed teams saw great promise in \Comet, and expressed interest in adoption.

\vspace{-0.2cm}
\section{Threats to Validity}
\label{sec:limitations}
\vspace{-0.0cm}

	Similar to past work on automated trace link recovery approaches~\cite{Guo:ICSE'17}, our work exhibits two main threats to validity. The first of these threats is to external validity. We utilize a limited number of systems to carry out our evaluation of \Comet, and thus it is possible that our results may not generalize to other systems. However, the systems utilized in our evaluation have been widely used in past work, and are of varying sizes and domains. We also examine one industrial grade open source project developed by our partner.
	
	The second threat to validity that affects experimental evaluation concerns construct validity, and more specifically, the accuracy of the ground truth for our subjects, and our implementation of the DL approach by Guo \etal~\cite{Guo:ICSE'17}. We cannot guarantee that the ground truth links for all of subjects are perfectly accurate. However, the ground truth sets for the CoEST datasets have been accepted by several pieces of prior work~\cite{Antoniol:e,Cleland-Huang:TSE'03,Poshyvanyk:TEFSE'11,Gethers:ICSM'11}. The ground truth for LibEST was derived by a team of the authors, and was validated with the help of industrial developers working on the project (see Sec. \ref{subsec:exp-context}). We re-implemented Guo \etal's DL approach closely following the details of the paper, although a full replication was not possible due to previous use of  proprietary industrial dataset. We will release our code~\cite{appendix} for this approach to aid in reproducibility. Another potential threat to validity is that our simulation of developer feedback in answering RQ$_2$ is not representative of real feedback. Due to constraints on developers time, we could not use real feedback, however, we believe simulating a small number of links using the ground truth, complete with error rates, represents a reasonable approximation.

\vspace{-0.2cm}
\section{Conclusion \& Future Work}
\label{sec:conclusion}
\vspace{-0.0cm}

We have presented \Comet, which takes a probabilistic view of the traceability problem and models the existence of traceability links using a Hierarchical Bayesian Network. We have shown that \Comet performs more consistently than IR/ML techniques, and has promising industrial applicability. We plan to adapt \Comet to new information sources, investigate tailoring \Comets analysis to infer security-related links, and further deploy the \Comet plugin with our industrial collaborators for feedback.
\vspace{-0.2cm}

%\vspace{-0.2cm}
\begin{acks}

This work is supported in part by Cisco Systems Inc. and the NSF CCF-1927679 grant. Any opinions, findings, and conclusions expressed herein are the authors' and do not necessarily reflect those of the sponsors.

\end{acks}

\balance
\bibliography{references}
\bibliographystyle{ACM-Reference-Format}

\end{document}